\documentclass[12pt]{article}
\usepackage[utf8]{inputenc}

\usepackage{amsmath}
\usepackage{amsthm}
\usepackage{amsfonts}
\usepackage{amssymb}
\usepackage{graphicx}
\usepackage{mathtools}
\usepackage{natbib}
\usepackage{url}
\usepackage{bm}

\usepackage{geometry}
\usepackage[usenames]{color}
\usepackage{hyperref}
\hypersetup{
  linkcolor  = blue,
  citecolor  = blue,
  urlcolor   = blue,
  colorlinks = true,
} 

\newcommand{\R}{\mathbb{R}}
\newcommand{\Y}{\mathcal{Y}}
\newcommand{\X}{\mathcal{X}}
\newcommand{\Hcal}{\mathcal{H}}

\newcommand{\Hbf}{\textbf{H}}

\newcommand{\V}{\textbf{V}}

\newcommand{\y}{\textbf{y}}
\newcommand{\x}{\textbf{x}}
\newcommand{\h}{\textbf{h}}

\newcommand{\s}{\textbf{s}}
\newcommand{\tbf}{\textbf{t}}

\newcommand{\lambdabf}{\boldsymbol{\lambda}}

\DeclareMathOperator{\E}{E}
\DeclareMathOperator{\Var}{Var}

\allowdisplaybreaks

\title{SEAM methodology for context-rich player matchup evaluations in baseball}
\author{Julia Wapner\thanks{jwapner2@illinois.edu}, 
David Dalpiaz\thanks{dalpiaz2@illinois.edu},
and Daniel J. Eck\thanks{dje13@illinois.edu} \\[1em]
Department of Statistics, University of Illinois Urbana-Champaign
}
\date{}

\begin{document}

\maketitle

\begin{abstract}
We develop the SEAM (synthetic estimated average matchup) method for describing batter versus pitcher matchups in baseball. We first estimate the distribution of balls put into play by a batter facing a pitcher, called the empirical spray chart distribution. Many individual matchups have a sample size that is too small to be reliable for use in predicting future outcomes. Synthetic versions of the batter and pitcher under consideration are constructed in order to alleviate these concerns. Weights governing how much influence these synthetic players have on the overall estimated spray chart distribution are constructed to minimize expected mean square error. We provide a Shiny web application that allows users to visualize and evaluate any batter-pitcher matchup that has occurred or could have occurred during the Statcast era (specifically 2017-present). This methodology and web application could be used to determine defensive alignments, lineup construction, or pitcher selection through estimation of spray densities based on any input matchup. One can access this web application at \begin{center}
\url{https://seam.stat.illinois.edu/}
\end{center} 
The computational speed with which the method calculates the spray densities allows the app to display the visualizations for any input almost instantly. Therefore, SEAM offers distributional interpretations of dependent matchup data which is computationally fast.
\end{abstract}


\section{Introduction}

Baseball has a rich statistical history dating back to the first box score created by Henry Chadwick in 1859. Throughout the years, fans of the game, academics, and professionals alike have made use of baseball statistics as a means of quantifying and comparing the ability of baseball players \citep{berry1999bridging, schwarz2004numbers, albert2006pitching, brown2008season, jensen2009hierarchical, jensen2009bayesball, piette2012estimating, baumer2015openwar, marchi2019analyzing}. Most statistics arrive from the box score, a simple summary of many discrete events that occurred throughout the course of a baseball game. Many additional ``advanced" baseball statistics combine these simple box score statistics and account for some contextual information including stadium and league effects \citep{marchi2019analyzing}. One can view the glossary on MLB.com (\url{https://www.mlb.com/glossary}) or explore comprehensive websites \citep{bref, fangraphs, bp} for descriptions on a wide array of such statistics. This vast effort to quantify the ability of baseball players through statistics has been increasingly embraced by baseball organizations due to a successful track record for statistics to explain how players' abilities translate to winning games \citep{lewis2004moneyball}.

The basic box score statistics which serve as the foundation for most baseball statistics do not represent the fundamental unit for which outcomes are realized. Baseball is a game with a pitcher on the mound (and a defense behind him) who first throws a ball to a batter. From here the ball is put into play or it is not, and at this moment box score statistics are tabulated. When we look closer at this chain of events we see that the pitcher who throws the ball possesses an arsenal of pitches whose movement, velocity, and release point are unique to that pitcher. The batter also has a unique set of traits including how hard they hit the ball (often referred to as exit velocity), and the horizontal and vertical angles in which they hit the ball (often referred to as spray and launch angles respectively). When a ball is put into play it has a final location that is relevant to, but not included in, the box score. As recently as 2015, this type of high-resolution data has been collected and made publicly available \citep{statcast}. Proprietary sources maintain similar high-resolution data dating back several years prior to 2015 \citep{BIS}.

This high resolution data has ushered in a new sophisticated class of techniques for finding and exploiting players' strengths and weaknesses. Use of these techniques have affected player performance and outcomes, which is reflected but noted in the boxscore. As an example, \cite{jensen2009bayesball} showed how this data could be used to better quantify defensive ability using Baseball Information Solutions data through 2007. One of the most prominent examples of the success of this high-resolution data has been the implementation of defensive shifting based on the spray chart \citep{pettispray, marchi2019analyzing}, a plot of locations a batter has hit the ball in the past. Through the early part of the 2022 MLB season, MLB had seen infield shifts on 37\% of pitches \citep{baccellieri2022infield}. This rate of deployment demonstrates the effectiveness of the strategy.

Even though the shift has a demonstrated record of success, the strategy of the shift can be justified by batters' empirical spray chart or pitchers' empirical spray chart allowed. These charts are compiled marginally across all matchups and they therefore do not account for the interaction effect when a particular batter faces a particular pitcher. However, spray charts for specific batter-pitcher matchups are not often able to be reliably constructed due to sparse or non-existent matchup data. 

In this article, we develop SEAM methodology which alleviates the challenge posed by sparse matchup data through a principled incorporation of batted-ball outcomes of players that are similar to the batter and pitcher under study. The SEAM method poses each batter-pitcher matchup as having a spray chart distribution which are 2-dimensional spatial distributions representing the potential batted-ball locations when a particular batter faces off against a particular pitcher. 
Estimation of the specific batter-pitcher matchup involves batted-ball locations from three sources: 1) those from the batter under study vs the pitcher under study; 2) those from the batter under study vs a synthetic version of the pitcher under study; 3) those from a synthetic version of the batter under study vs the pitcher under study. Here the batted-ball locations corresponding to synthetic versions of the players under study are taken across all other MLB players, the locations are weighted to reflect similarity to the player under study where similarity ranges from 0 (not considered at all) to 1 (a clone of the player under study up to the information used to compute the similarity score). These three components represent separate spray chart distributions and they are combined through a convex combination with weights reflecting overall similarity of the synthetic players to those under study where these weights collapse on the specific batter-pitcher empirical spray distribution when enough data is available to reliably estimate this distribution. 

Mathematical justification for the combination of these estimated spray distributions is provided. We also demonstrate that our SEAM method outperforms the empirical batter or pitcher spray chart distributions through a detailed validation study in which we hold out data from the 2021 season and determine how well $(1-\alpha)\times 100\%$ highest density regions include the batted-ball locations from the held out data. SEAM exhibits nominal marginal coverage, but is no better than the existing batter or pitcher spray charts. SEAM  outperforms batter or pitcher spray charts when we evaluate conditional coverage, indicating that our approach is better suited for predicting batted-ball locations. That being said, the SEAM confidence region is larger than those constructed from empirical batter or pitcher spray chart distributions. We then consider fixed-size confidence regions as a compromise between size and coverage. We find that the SEAM confidence region has higher conditional coverage than its competitors when we compare fixed-size confidence regions. The size of these fixed-size regions is chosen to reflect the proportion of a baseball field that fielders could cover. Therefore, the SEAM method exhibits an important advantage over its competitors.

\section{SEAM method}

\subsection{Spray chart distribution preliminaries}

In this article, we seek to provide a batter-pitcher specific spray chart density which is more accurate than spray chart densities estimated exclusively from either batter or pitcher data marginally. The densities that we estimate correspond to what we call a spray chart distribution which is a spatial distribution over a bounded subset $\Y \in \R^2$. The set $\Y$ contains plausible locations of batted balls from home plate. In our application we will let $(0,0) \in \Y$ denote the location of home plate. A spray chart distribution for a batter is a distribution $F$ over a bounded subset $\Y \in \R^2$. 

\subsection{Nonparametric estimation}

A spray chart distribution $F$ is not likely to have a known parametric form so we will consider nonparametric estimation techniques to estimate spray chart densities $f$. The particular nonparametric estimation methods that we use are chosen for their computational speed and theoretical properties. Let $(y_{1i},y_{2i}) \in \Y$, $i = 1, \ldots, n$, be the observed batted-ball locations, then we will estimate $f$ with a multivariate kernel density estimator
\begin{equation} \label{general}
  \hat f_\Hbf(\y) = \frac{1}{n|\Hbf|}\sum_{i=1}^{n} K\left(\Hbf^{-1}(\y_i - \y)\right),
  \qquad \y \in \Y,
\end{equation}
where $K$ is a multivariate kernel function and $\Hbf$ is a matrix of bandwidth parameters. Our implementation will estimate $f$ using variants of the \texttt{kde2d} and \texttt{kde2d.weighted} functions in R \citep{MASS, ggtern}. 
Therefore, we estimate $f$ using a bivariate nonparametric Gaussian kernel density estimator
\begin{equation} \label{spraydens}
  \hat f_\h(\y) =
    \frac{1}{n h_{y_1}h_{y_2}}\sum_{i=1}^{n} \phi\left(\frac{y_1 - y_{1i}}{h_{y_1}}\right)
      \phi\left(\frac{y_2 - y_{2i}}{h_{y_2}}\right),
\end{equation}
where $\y = (y_1, y_2) \in \Y$, $\phi$ is a standard Gaussian density, $\h \in \R^2$ is a bandwidth parameter so that the matrix $\Hbf$ in \eqref{general} is $\Hbf = \text{diag}(\h)$, and $\Hbf$ is chosen according to the default bandwidth selection procedures within the \texttt{kde2d} and \texttt{kde2d.weighted} functions. The estimated spray chart density function $\hat f_\h$ is a smoothed surface overlaying a spray chart. Our visualization of the spray chart distribution will be along $n_g$ common grid points $g_1,\ldots,g_{n_g}$ for all matchups under study. Commonality of grid points allows for straightforward comparisons of spray chart distributions in practice.

We extend this framework to spray chart distributions that are conditional on several characteristics for pitchers $\x_p$ and batters $\x_b$, where $\x = (\x_p',\x_b')' \in \X$, and $\X$ is assumed to be bounded. Denote the conditional spray chart distribution as $F(\y|\x)$ for $\x \in \X$ and $\y \in \Y$. The conditional spray chart density function corresponding to $F(\y|\x)$ will be denoted as $f(\y|\x)$ for all $\x \in \X$ and $\y \in \Y$. Thus, $f(\y|\x)$ is a nonparametric regression model that we will again estimate with a bivariate nonparametric Gaussian kernel density estimator
\begin{equation} \label{spraydens-cov}
  \hat f_\h(\y|\x) =
    \frac{1}{n h_{y_1}h_{y_2}}\sum_{i=1}^{n} \phi\left(\frac{y_1 - y_{1i}}{h_{y_1}}\right)
      \phi\left(\frac{y_2 - y_{2i}}{h_{y_2}}\right),
\end{equation}
where the sample of batted-ball locations $(y_{1i},y_{2i}) \in \Y$, $i = 1, \ldots, n$ are now conditional on $\x \in \X$.

\subsection{Similarity scores}

The estimator \eqref{spraydens-cov} is often not feasible for $f(\cdot|\textbf{x})$ in practice since there is not enough individual matchup data. To address this challenge we will incorporate batted-ball data from other matchups involving one of the batter or pitcher under study where the batted-ball locations from such matchups are weighted by their similarity to the players in the matchup under study. Similarity will be assessed via similarity scores which are motivated by \cite{james1994politics} and \cite{PECOTA}.


We now describe our procedure for determining similarity scores. We will suppose that there are $J$ pitchers who allowed a ball in play against the batter under study and $K$ batters who put a ball in play against the pitcher under study. We will suppose that the pitcher in the matchup under study throws $n_{\text{type}}$ different types of pitches. We will let $\x_{p,t}$ be the pitcher covariates for pitch type $t = 1,\ldots,n_{\text{type}}$. Similarly, let $\x_{b,t}$ be the batter covariates when facing pitch type $t = 1,\ldots,n_{\text{type}}$. The covariates in $\x_{p,t}$ and $\x_{b,t}$ are averages across the pitch-by-pitch realizations. We will denote $d_p$ and $d_b$ as the dimensions of $\x_{p,t}$ and $\x_{b,t}$ respectively. For a pitch type $t$, the similarity score of pitcher $j_1$ to pitcher $j_2$ is defined as 
$$
  s(\x_{p,j_1,t}, \x_{p,j_2,t}) = \exp(-\|\x_{p,j_1,t}-\x_{p,j_2,t}\|_{\V_{p,t}}),
$$ 
where $\x_{p,j_1,t}$ and $\x_{p,j_2,t}$ are, respectively, the underlying pitch characteristics for pitcher $j_1$ and $j_2$,
\begin{equation} \label{Vpt}
   \|\x_{p,j_1,t}-\x_{p,j_2,t}\|_{\V_{p,t}}
     = \left((\x_{p,j_1,t}-\x_{p,j_2,t})'\V_{p,t}(\x_{p,j_1,t}-\x_{p,j_2,t})\right)^{1/d_p},
\end{equation}
and $\V_{p,t}$ is a diagonal weight matrix that is chosen to scale the pitch characteristics and give preference to pitch characteristics that are chosen to have higher influence on the spray chart distribution under study. Similarity scores of the form $s(\x_{p,j_1,t}, \x_{p,j_2,t})$ have desirable theoretical properties that are explained in the Appendix and, in practice, they guard against downplaying the effect of the players under study. Users of our web application have some control of the entries of $\V_{p,t}$ in \eqref{Vpt} by adjusting the pitcher slider. Similarity scores between batters $k_1$ and $k_2$ are defined in a similar manner and are denoted as $s(\x_{b,k_1,t}, \x_{b,k_2,t})$.

Implicit in this construction is the assumption that the collected pitcher and batter characteristics are an exhaustive set of inputs to properly estimate the spray chart distribution. Therefore, we are assuming that $f$ is conditional on $\x_{b,t}, \x_{p,t}, \rho_t$, for $t = 1,\ldots,n_{\text{type}}$, where $\rho_t$ is the proportion of time that a ball in play yielded by the pitcher in the matchup under study corresponds to pitch type $t$. We therefore represent $f(y)$ as $\sum_t \rho_t f(y|\x_t)$, where $\x_t = (\x_{p,t}', \x_{b,t}')'$.

\subsection{The SEAM density function}

We now describe the construction and estimation of the SEAM density function. The SEAM density function is a linear combination of three density functions, the spray chart density $f(\cdot|\textbf{x})$, and two ``synthetic" player density functions, the synthetic pitcher density function $f_{\text{sp}}(\cdot|\textbf{x})$ and the synthetic batter density function $f_{\text{sb}}(\cdot|\textbf{x})$. Both $f_{\text{sp}}(\cdot|\textbf{x})$ and $f_{\text{sb}}(\cdot|\textbf{x})$ are weighted averages of spray chart densities corresponding to matchups in which the batter (pitcher) under study faces every other pitcher (batter). Attention is restricted to matchups that yielded a ball in play. We now describe these synthetic densities in more detail, describe their estimation, and present an estimator of the SEAM density function.

Without loss of generality, let $\x_{p,t}$ be the characteristics for pitch type $t$ thrown by the pitcher under study, let $\x_{b,t}$ be the characteristics for the batter under study. We then line up the pitcher characteristics for all of the pitchers in the donor pool, $\x_{p,j,t}$, $j = 1,...,J$. Now obtain the similarity scores $s_{p,j,t} = s(\x_{p,t},\x_{p,j,t})$ and then construct the weights $w_{p,j,t} = s_{p,j,t} / \sum_{l=1}^{J}s_{p,l,t}$, for $j = 1,...,J$. For pitch type $t$, the spray chart density for a batter facing the synthetic pitcher is
\begin{equation} \label{synth-pitch-t}
  f_{\text{sp}, t}(\y|\x_{b,t}) = \sum_{j=1}^J w_{p,j,t}f(\y|\x_{p,j,t},\x_{b,t}).
\end{equation}
The spray chart density for a batter facing the synthetic pitcher is then
\begin{equation} \label{synth-pitch}
  f_{\text{sp}}(\y) = \sum_{t=1}^{n_{\text{type}}} \rho_t f_{\text{sp},t}(\y|\x_{b,t}).
\end{equation}
The conditioning on $\x_{b,t}, \x_{p,j,t}, \rho_t$, for $t = 1,\ldots,n_{\text{type}}$ and $j = 1,\ldots,J$ is suppressed in the density $f_{\text{sp}}(\y)$.

Similarly, we describe the synthetic spray chart density for the synthetic batter facing the pitcher under study. For pitch type $t$, we line up the batter characteristics for all of the available batters that faced pitch type $t$ thrown by the pitcher under study, $\x_{b,k,t}$, $k = 1,...,K$.  We obtain the similarity scores $s_{b,k,t} = s(\x_{b,t},\x_{b,k,t})$ and then construct the weights $w_{b,k,t} = s_{b,k,t} / \sum_{l=1}^{K}s_{b,l,t}$, for $k = 1,...,K$. For pitch type $t$, the spray chart density for a pitcher facing the synthetic batter is
\begin{equation} \label{synth-bat-t}
  f_{\text{sb},t}(\y|\x_{p,t}) = \sum_{k=1}^K w_{b,k,t}f(\y|\x_{p,t},\x_{b,k,t}).
\end{equation}
The spray chart density for the synthetic batter facing the pitcher under study is then
\begin{equation} \label{synth-bat}
  f_{\text{sb}}(\y) = \sum_{t=1}^{n_{\text{type}}} \rho_t f_{\text{sb},t}(\y|\x_{p,t}).
\end{equation}
The conditioning on $\x_{p,t}, \x_{b,k,t}, \rho_t$, for $t = 1,\ldots,n_{\text{type}}$ and $k = 1,\ldots,K$ is suppressed in the density $f_{\text{sb}}(\y)$.

We then estimate \eqref{synth-pitch-t} and \eqref{synth-bat-t} with
\begin{equation} \label{synth-est-t}
  \hat f_{\text{sp},t}(\y|\x_{b,t}) = \sum_{j=1}^J w_{p,j,t}\hat f_{\h_{p,j,t}}(\y|\x_{p,j,t},\x_{b,t}),
  \qquad
  \hat f_{\text{sb}, t}(\y|\x_{p,t}) = \sum_{k=1}^K w_{b,k,t}\hat f_{\h_{b,k,t}}(\y|\x_{p,t},\x_{b,k,t}),
\end{equation}
where, for pitch type $t$, we let $n_{p,j,t}$ denote the matchup sample size of pitcher $j$ versus the batter under study, $n_{b,k,t}$ denote the matchup sample size of the pitcher under study versus batter $k$, and $\h_{p,j,t}$ and $\h_{b,k,t}$ are bandwidth parameters. We estimate the densities in \eqref{synth-pitch} and \eqref{synth-bat} with,
\begin{equation} \label{synth-est}
  \hat f_{\text{sp}}(\y) =  \sum_{t=1}^{n_{\text{type}}} \rho_t \hat f_{\text{sp}, t}(\y|\x_{b,t}),
  \qquad
  \hat f_{\text{sb}}(\y) = \sum_{t=1}^{n_{\text{type}}} \rho_t
    \hat f_{\text{sb}, t}(\y|\x_{p,t}).
\end{equation}

The estimators \eqref{synth-est} are obviously biased estimators for $f$. However, they have the potential to reduce MSE. One obvious case is when, for all
$t = 1,\ldots,n_{\text{type}}$, there exists weights
$w_{p,j,t}, w_{b,k} \approx 1$ and
$n_{p,j,t}, n_{b,k,t} > n$.
In such settings, $f_{\text{sp}}(\y)$ and $f_{\text{sb}}(\y)$ have minimal bias when estimating $f$ and can be more efficient than $\hat f_h$. Another obvious case is when the batter has never faced the pitcher so that no data is available to estimate $f$ directly, although that does not guarantee that the estimators \eqref{synth-est} are good estimators for $f$. Our implementation will estimate $f$ with
\begin{equation} \label{sd-implem}
  \hat{f}_{\lambdabf}(\y) = \lambda \hat f_\h(\y)
    + \lambda_p \hat f_{\text{sp}}(\y)
    + \lambda_b \hat f_{\text{sb}}(\y)
\end{equation}
where $\lambda,\lambda_p$, $\lambda_b$ form a convex combination. The conditioning on $\x_{p,j,t}, \x_{b,k,t}, \rho_t$, for $t = 1,\ldots,n_{\text{type}}$ and $k = 1,\ldots,K$ is suppressed in the density $\hat{f}_{\lambdabf}(\y)$.
Our implementation will estimate the elements of $\lambdabf$ as
$$
  \lambda = \frac{\sqrt{n}}{\sqrt{n} + \sqrt{n_p} + \sqrt{n_b}}, \qquad
  \lambda_p = \frac{\sqrt{n_p}}{\sqrt{n} + \sqrt{n_p} + \sqrt{n_b}}, \qquad
  \lambda_b = \frac{\sqrt{n_b}}{\sqrt{n} + \sqrt{n_p} + \sqrt{n_b}},
$$
where $n_p = \sum_t\rho_t\sum_{j=1}^J s_{p,j,t}^2n_{p,j,t}$ and
$n_b = \sum_t\rho_t\sum_{k=1}^K s_{b,k,t}^2n_{b,k,t}$.  These choices arise as a balance between the natural bias that exists in our synthetic player construction and the inherent estimation variation. 
It is reasonable to assume that $n_{p,j,t} = O(n)$ and $n_{b,k,t} = O(n)$ for all $j = 1,\ldots,J$, all $k = 1,\ldots,K$, and all $t = 1,\ldots,n_{\text{type}}$. It is also reasonable to assume that $n$ will be too small to be of much use, hence the reason why $n_p$ and $n_b$ are aggregated with respect to similarity scores instead of weights that form a convex combination. However, in the event that $n$ is large enough to provide reliable estimation of $f(\y|\x)$ with $\hat f_h(\y|\x)$, then $n$ dominates $n_p$ and $n_b$. Formal technical justification for selecting $\lambdabf$ is given in the Appendix. In the Appendix we argue that our choices of $\lambdabf$ lead to the estimator \eqref{sd-implem} having a lower MSE than the estimator \eqref{spraydens}.

\section{Validation}

We evaluate the performance of our SEAM method through a validation approach in which we train the SEAM method on data through the 2020 season and then test the coverage of highest density regions with 2021 data. We define coverage as the proportion of batted balls that land in a location on the field that is within a level $\alpha$ highest density region. We consider both marginal and conditional coverage. Marginal coverage is assessed across all 2021 batted balls. Conditional coverage is assessed for 2021 batter-pitcher matchups that yielded 10 or more balls in play. We compute the proportion of individual batter-pitcher conditional coverages that at or above the nominal level, which we call conditional coverage success rate.

\begin{table}
\begin{center}
\begin{tabular}{l|rrr ||rrr }
  & \multicolumn{3}{c||}{Marginal coverage} & \multicolumn{3}{|c}{Conditional coverage success rate} \\
  \hline
level  & 0.5 & 0.75 & 0.9 & 0.5 & 0.75 & 0.90 \\
\hline
SEAM & 0.559 & 0.811 & 0.953 & 0.579 (0.035) & 0.641 (0.034) & 0.779 (0.030) \\
batter & 0.547 & 0.793 & 0.938 & 0.513 (0.036) & 0.574 (0.035) & 0.662 (0.034) \\
pitcher & 0.541 & 0.788 & 0.935 & 0.549 (0.036) & 0.595 (0.035) & 0.713 (0.032) \\
\end{tabular}		
\end{center}
\caption{Marginal and conditional coverage properties for SEAM, the empirical batter spray distribution, and the empirical pitcher spray distribution (standard errors in parentheses). The conditional coverage success rate is the proportion of individual batter-pitcher conditional coverages that at or above the nominal level. }
\label{Tab:coverage}
\end{table}

We compare the coverage properties of our SEAM method to empirical batter spray charts and empirical pitcher spray charts which are estimated marginally across all pitchers and batters respectively using similar nonparametric estimation techniques. The results are included in Table~\ref{Tab:coverage}. We see that the highest density regions for each method achieve nominal marginal coverage. This is especially encouraging for SEAM since it is expected that the empirical batter and pitcher spray charts should obtain marginal coverage by their construction, provided that the players under study did not dramatically change in a material way. We also see that the conditional coverage of the SEAM method is noticeably and usefully higher than the other methods. 

However, the highest density region for SEAM is slightly larger than its competitors. As an example, the level $\alpha = 0.10$ highest density region for SEAM includes 34\% of grids forming the rectangle that contains the playing filed while similar highest density regions corresponding to the empirical batter and pitcher spray distributions include, respectively, 33.4\% and 33\% of such grids. 

\begin{figure}[t!]
\begin{center}
\includegraphics[width=0.80\textwidth]{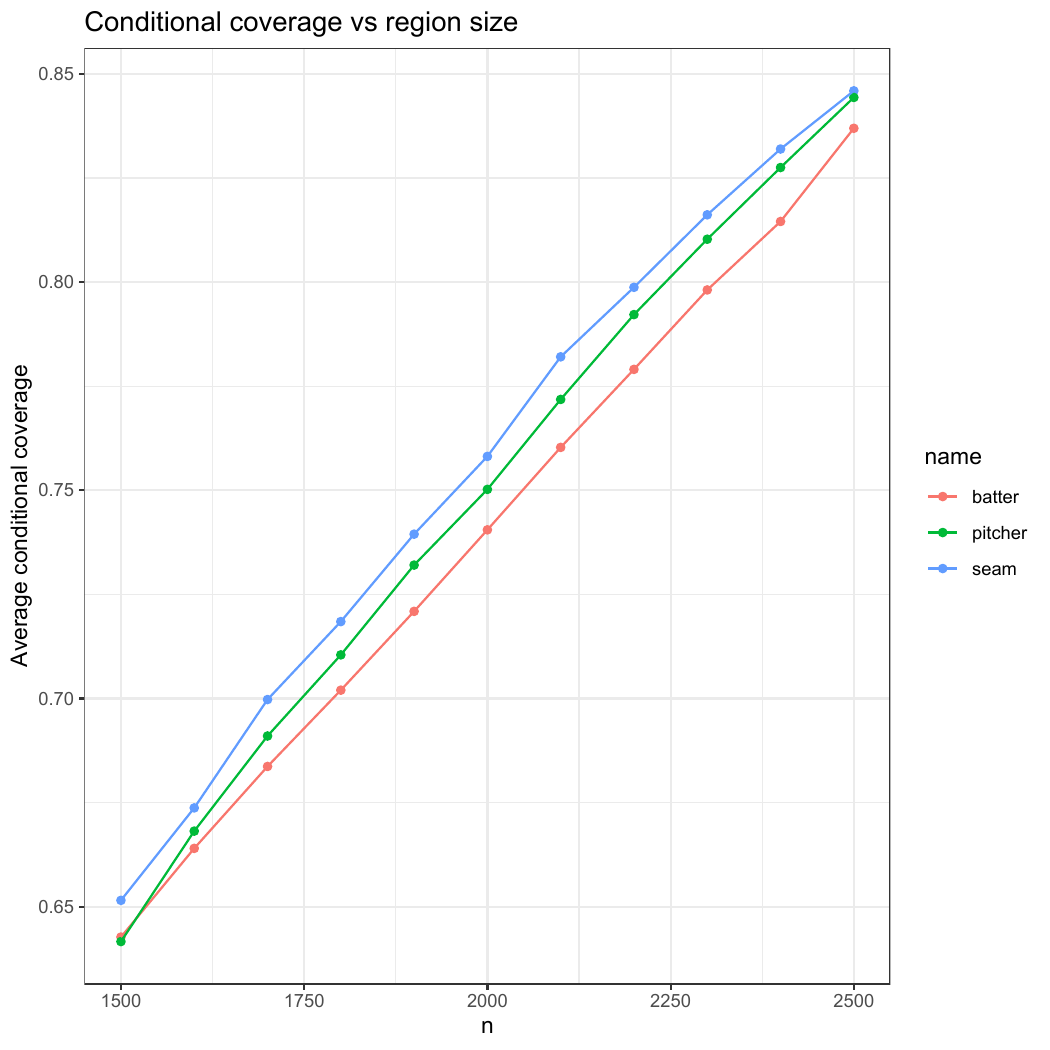}	
\end{center}	
\caption{Average conditional coverage of fixed-size confidence regions for SEAM, the empirical batter spray distribution, and the empirical pitcher spray distribution. The fixed-size regions are constructed from the $n$ most probable grids according to each method. }
\label{Fig:fixed-size}
\end{figure}

We will now consider a different type of confidence region which will balance these conflicting conclusions with size and coverage. This region is constructed using the $n$ grids containing the highest probable batted ball locations as estimated by each method. Calculating average conditional coverages for fixed-size regions is an important consideration for fielding placement due to the physical constraint on how much of the baseball field defenders can reasonably cover. We considered [1500, 2500] as a range for the number of grids. This range corresponds to fixed-size regions that cover approximately [50\%,\;80\%] of fair territory of a baseball field that is 115,000 square feet. 
We then assess the average conditional coverage of these fixed-size regions. What we find is that the SEAM fixed-size region has higher average conditional coverage than its competitors which ignore matchup considerations. This is despite the fact that the cumulative probability mass of the SEAM fixed-size region is lower than its competitors. The results are in Table~\ref{Tab:fixed-size}, and Figure~\ref{Fig:fixed-size}. 


\begin{table}
\begin{center}
\begin{tabular}{c|ccc}
$n$ & SEAM & batter & pitcher\\
\hline
1500 & 0.652 & 0.643 & 0.642 \\
1600 & 0.674 & 0.664 & 0.668 \\
1700 & 0.700 & 0.684 & 0.691 \\
1800 & 0.718 & 0.702 & 0.710 \\
1900 & 0.739 & 0.721 & 0.732 \\
2000 & 0.758 & 0.741 & 0.750 \\
2100 & 0.782 & 0.760 & 0.772 \\
2200 & 0.799 & 0.779 & 0.792 \\
2300 & 0.816 & 0.798 & 0.810 \\
2400 & 0.832 & 0.815 & 0.828 \\
2500 & 0.846 & 0.837 & 0.844 
\end{tabular}
\end{center}
\caption{Average conditional coverages of fixed-size confidence regions for SEAM, the empirical batter spray distribution, and the empirical pitcher spray distribution. The fixed-size regions are constructed from the $n$ most probable grids according to each method.}
\label{Tab:fixed-size}
\end{table}

\section{Data considerations}

Our methodology will consider the following variables comprising $\x_{p,t}$: velocity, spin rate, horizontal break, horizontal release angle, horizontal release point, vertical break, vertical release angle, vertical release point, and extension. Averages of these variables are taken across each pitcher-pitch type-year combination. Our methodology will consider the following variables comprising $\x_{b,t}$: exit velocity, launch angle, left\%, center\%, and right\%. Averages of these variables are taken across each batter-pitch type-year combination. We separate for year to account for subtle season-to-season shifts in pitcher and batter characteristics. One should note that these variables will not allow us to measure the complete talent profile of baseball players. 
Skills such as speed and eye at the plate will not be fully captured by our methodology.

Data for our web application was acquired via Statcast \citep{statcast}. Our data set contains every pitch that was thrown since the start of the 2017 season. We ignored Statcast data pre-2017 due to data integrity issues. A few preprocessing steps are involved:
\begin{itemize}
    \item Pitches classified as Eephus, Knuckleball, and Screwball are removed since these pitch types are rare.
    \item Pitches classified as Knuckle-Curve are renamed to Curveball.
    \item Pitches classified as Forkball are renamed to Splitter.
    \item Pitch launch angles are calculated using rudimentary kinematics:
        \begin{itemize}
            \item $launch_h = \arctan(\frac{vx_r}{vy_r})$
            \item $launch_v = \arctan\left(\frac{vz_r}{\sqrt{vx_r^2 + vy_r^2}}\right)$
        \end{itemize}
        where $vx_r$, $vy_r$, $vz_r$ are, respectively, the $x$, $y$, $z$ components of release velocity.
    \item Batted ball locations are adjusted to make home plate the origin $(0,0)$ \citep{petti2017research}.
    \item Spray angle \citep{petti2017research} is calculated from the $x$ and $y$ coordinates of the batted ball locations (what we previously called $y_1$ and $y_2$).
    \item Data was limited to regular season batter-pitcher matchups. 
    \item Sacrifice hits and sacrifice flies are removed from consideration.
\end{itemize}

Pitchers and batters are aggregated on a season, handedness, and pitch type basis. 

\section{Interactive web application}

\begin{figure}
\centering
    \includegraphics[width=5.5in, height=3in]{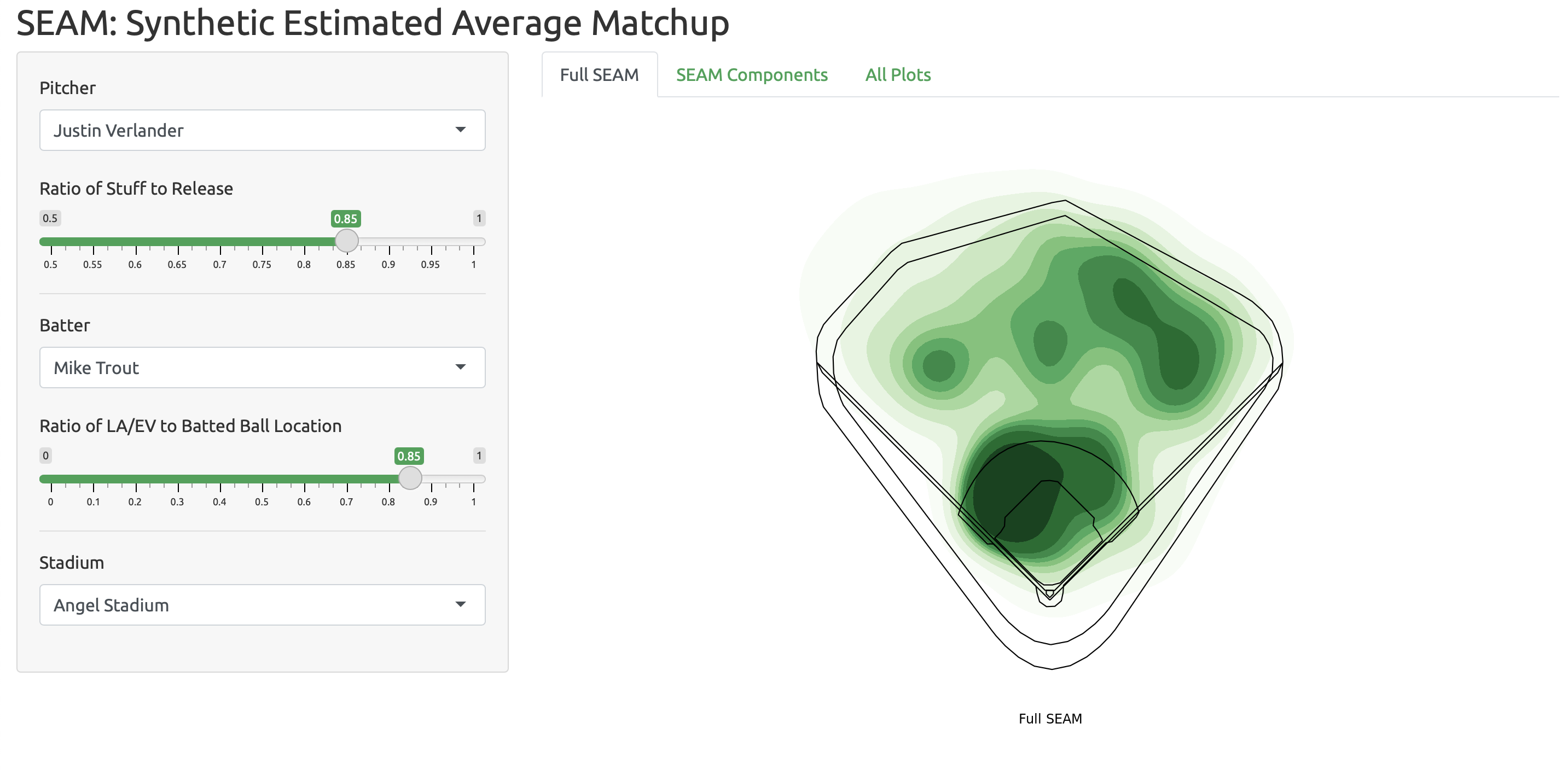}
    \caption{The layout of the web application upon submission.}
    \label{layout}
\end{figure}

\begin{figure}
\centering
    \includegraphics[width=0.90\textwidth]{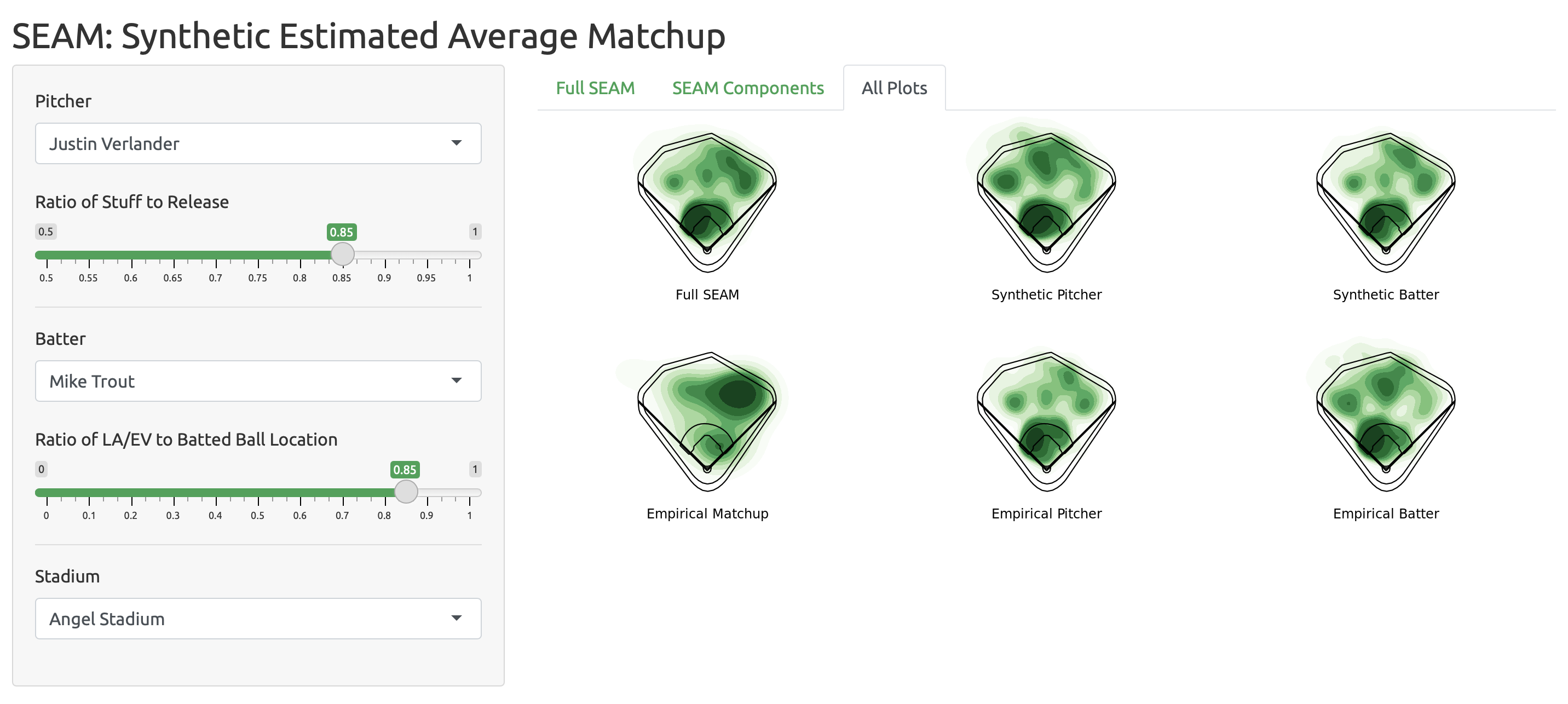}
    \caption{Spray chart distributions constructed by our web application.  This example
    corresponds to the spray chart distribution when batter Mike Trout faces
    pitcher Justin Verlander.
    The top-left panel is the complete synthetic spray chart for
      the batter-pitcher matchup.
    The top-center panel is the synthetic pitcher's spray chart distribution
      versus the real batter.
    The top-right panel is the synthetic batter's spray chart distribution
      versus the real pitcher. 
    The bottom-left panel is the traditional batter-pitcher spray chart
      distribution, with no consideration of similar players.   
    The bottom-center panel is the real pitcher's spray chart distribution versus all batters of the same handedness as the matchup under study.
    The bottom-right panel is the real batter's spray chart distribution versus all pitchers of the same handedness as the matchup under study.
}
    \label{spraydists}
\end{figure}

In this section we present a snapshot of what our Shiny web application implementing SEAM methodology offers users. The web application is available at 
\begin{center}
\url{https://seam.stat.illinois.edu/} 
\end{center}
The default matchup in the web application pairs the 2019 American League Cy Young winner Justin Verlander against the 2019 American League MVP Mike Trout, two of the greatest players of all-time \citep{yan2022comparing}. Note that our matchup will assess the outcomes currently, rather than during their award winning seasons. The layout includes a sidebar with five filters: two dropdowns for batter/pitcher selection, two sliders for metric adjustment, and a dropdown to select the stadium appearance. A snapshot of the appearance of our visualization is depicted in Figures \ref{layout} and \ref{spraydists}.

The pitcher slider allows users to determine the relative importance of ``stuff'', a colloquial term for pitch quality, versus release information. Stuff includes velocity, spin rate, and movement. Release includes release angles and release point. The batter slider allows users to determine the relative importance of launch conditions versus batted ball locations. Launch conditions includes exit velocity and launch angle. Location includes pull\%, middle\%, oppo\% (the percentage of batted balls place into the corresponding thirds of a baseball field). The default setting of the pitcher slider favors stuff over release information. The logic for this is quality of pitches being more representative of ability than release point. The default setting of the batter slider favors quality of contact over batted ball tendencies which appears to bias the synthetic batter's spray chart away from that of the batter under consideration. That being said, the batted ball tendencies are recorded as percentages of balls hit to six large grids on the baseball field, ignoring the quality, trajectory, and exact location of the batted ball. Thus, the quality of contact forms a more complete representation of a batter's skill than tendency.

Note that the same slider weights are applied to each pitch type so that $V_{p,t} = V_p$ in \eqref{Vpt}, the same holds for batter characteristics. Additionally, our implementation calculates $\rho_t$ as the marginal proportion of pitch types computed across all pitches thrown by the pitcher that yielded a ball in play. This design construction is appropriate for the descriptive nature of the analyses presented. For more prescriptive uses, one could adapt more flexible choices of $\rho_t$ to incorporate contextual information known about a batter-pitcher matchup, possibly in real time. Further note that pitch proportions are not considered in our similarity score constructions, therefore we are not accounting for the batter-pitcher meta game.

\section{Discussion}

The primary contribution of this work is the development of SEAM methodology in which a synthetic spray chart density function $f_{\lambdabf}(\y)$ is estimated. In our presented context of batter-pitcher matchups, this estimated density function is a weighted average of $\hat f_\h(\y|\x)$, $\hat f_{\text{sp}}(\y)$, and $\hat f_{\text{sb}}(\y)$, where these weights are chosen to minimize MSE under an assumed smooth function space. The synthetic players are constructed to best mimic the players under study. It was demonstrated that our SEAM method provided more accurate predictions of future batted-ball locations than spray chart distributions computed from either the batter or pitcher in isolation. Our method of synthetic player construction is generalizable to other settings in baseball as well as other sports. We are not the first to incorporate additional players into an analysis via similarity scores with the understanding that doing so improves estimation performance. The PECOTA prediction methodology \citep{PECOTA} tries to forecast the ability of players using aggregate estimates obtained from other similar players. To the best of our knowledge, we are the first to base similarity scores exclusively on Statcast data which we believe provides a truer notion of talent similarity.

We also developed an interactive web application which implements and provides visualizations of our SEAM methodology. This web application is fast, providing users with visual measures of batter-pitcher matchups nearly instantaneously. Our methodology displays a vast improvement upon empirical batter or pitcher spray charts which are computed marginally \citep{pettispray, marchi2019analyzing}. Additionally, matchup-specific spray charts are often not informative due to a lack of data. Our synthetic player construction alleviates this problem.

As previously mentioned, the visualizations provided by our web application can help coaches position their fielders effectively. While a traditional spray chart may be useful in aggregate, building a custom spray chart to reflect a specific batter-pitcher matchup will yield more accurate results on a plate appearance by plate appearance level. This synthetically created spray chart will give the user an expected distribution of batted balls for the batter-pitcher matchup based on a combination of the distribution of similar batters against the pitcher, the distribution of similar pitchers against the batter, and the distribution of any observations of the pitcher vs batter since 2017. 

The default matchup between Justin Verlander and Mike Trout provides a good example of how to interpret our SEAM spray chart distribution estiamtes. Trout seems to be a pull-heavy hitter in general according to his traditional marginal spray chart. When facing pitchers similar to Verlander, he seems to push the ball the opposite way. This may be explained by Verlander's high velocity fastball. In general, batters have a hard time ``getting around'' (pulling) an upper-90's fastball, so they end up hitting the ball to the opposite field. Given this spray chart distribution, a coach may position the shortstop more towards third base, the second baseman more up the middle, and the first baseman more towards second base. This will protect against Trout's usual habit of pulling the ball, and also put the first baseman in a position to cover the opposite field soft ground ball. If this decision were made just by Trout's traditional chart, the first baseman might not have been moved to cover ground balls through the right side.



\bibliographystyle{plainnat}
\bibliography{spray}

\section*{Appendix: Justification for our choice of $\lambdabf$}

We now motivate $\lambdabf$ theoretically. We first assume some additional structure on the space of functions that $f(\cdot|\cdot)$ belongs to in order to facilitate our motivation. The best batters in baseball are good at hitting the ball with general intent but batted ball locations will still exhibit variation. Therefore we expect spray chart densities to be smooth and lacking sharp peaks. It is reasonable to assume that $f(\cdot|\cdot)$ belongs to a multivariate H{\"o}lder class of densities which we will denote by $H(\beta,L)$. The space $H(\beta,L)$ is the set of functions $f(\y|\x)$ such that
\begin{align*}
|D_{\y}^\s f(\y|\x) - D_{\y}^\s f(\y'|\x)| &\leq L_\x\|\y - \y'\|^{\beta - |\s|}, \\
|D_{\x}^\tbf f(\y|\x) - D_{\x}^\tbf f(\y|\x')| &\leq L_\y\|\x - \x'\|^{\beta - |\tbf|},
\end{align*}
for all $\y,\y' \in \Y$, all $\x,\x' \in \X$, and all $\s$ such that $|\s| = \beta - 1$ where
$D_{\y}^\s = \partial^{s_1 + s_2}/\partial y_1^{s_1} \partial y_2^{s_2}$,
$D_{\x}^\tbf = \partial^{t_1 + \cdots + t_p}/\partial x_1^{t_1} \cdots \partial x_p^{t_p}$ and $L_\x \leq L$ for all $\x \in \X$ and $L_\y \leq L$ for all $\y \in \Y$.
We will assume the following regularity conditions for our spray chart distributions and kernel functions:

\begin{itemize}
\item[A1.] The density $f$ is square integrable, twice continuously differentiable, and all the second order partial derivatives are square integrable. We will suppose that $\beta = 2$ in $H(\beta,L)$.
\item[A2.] The kernel $K$ is a spherically symmetric and bounded pdf with finite second moment and square integrable.
\item[A3.] $\Hbf = \Hbf_n$ is a deterministic sequence of positive definite symmetric matrices such that, $n\det(\Hbf) \to \infty$ when $n \to \infty$ and $\Hbf \to 0$ elementwise.
\end{itemize}

Condition A2 holds for the multivariate Gaussian kernel function that we use in our implementation. We will let $\Hbf$ be a matrix of bandwidth parameters that has diagonal elements $\h$, in our implementation $\Hbf = \text{diag}(\h)$. We will assume that $\h = \h_t$, the bandwidth parameters for the batter-pitcher matchup are the same across pitch types. We will use the following notation: $R_{\x}(f) = \int f(\y|\x)^2 d\y$, $\mu_2(K) = \int u^2K(u)du$, and $\Hcal_f(\y|\x)$ is the Hessian matrix respect to $f(\y|\x)$ where derivatives are taken with respect to $\y$. Assume that pitch outcomes are independent across at bats and that $n_{p,j,t} = O(n)$, $n_{b,k,t} = O(n)$ and $\h_{p,j,t} = O(\h)$, $\h_{b,k,t} = O(\h)$ for all $j = 1, \ldots, J$, $k = 1, \ldots, K$, $t = 1,\ldots,n_{\text{type}}$. Standard results from nonparametric estimation theory give
$$
  \E(\hat f_\h(\y|\x)) - f(\y|\x) = \frac{\mu_2(K)\h'\text{diag}(\Hcal_f(\y|\x))\h}{2}
    + o(\|\h\|^2),
$$
and
$$
 \Var(\hat f_\h(\y|\x)) = \frac{R_{\x}(f)f(\y|\x)}{n\det(\Hbf)} + O\left(\frac{1}{n}\right).
$$
With the specification that $\beta = 2$ in Condition A1 we have that
$
   f(\y|\x) - L\|\x-\x'\|^2 \leq f(\y|\x') \leq f(\y|\x) + L\|\x-\x'\|^2.
$
This result implies that
\begin{align*}
  R_{\x'}(f) - R_\x(f) &= \int (f(\y|\x')^2 - f(\y|\x)^2) d\y
     = \int (f(\y|\x') - f(\y|\x))(f(\y|\x') + f(\y|\x)) d\y \\
  &\leq L\|\x'-\x\|^2 \int(f(\y|\x') + f(\y|\x)) d\y
    = 2L\|\x'-\x\|^2,
\end{align*}
and
$
   R_\x(f) - 2L\|\x-\x'\|^2 \leq R_{\x'}(f) \leq R_\x(f) + 2L\|\x-\x'\|^2.
$

We now have enough structure to estimate the MSE of \eqref{spraydens} and \eqref{sd-implem}.
Our multivariate H{\"o}lder class specifications yield,
\begin{align*}
  &\E(\hat f_{\lambdabf}(\y)) = \lambda \E \hat f_\h(\y)
    + \lambda_p \E \hat f_{\text{sp}}(\y)
    + \lambda_b \E \hat f_{\text{sb}}(\y) \\
  &\qquad= \lambda f(\y)
    + \lambda \sum_t\rho_t\frac{\mu_2(K)\h'\text{diag}(\Hcal_f(\y|\x_t))\h}{2}
    + \lambda_p\sum_t\rho_t \sum_{j=1}^J w_{p,j,t} \E \hat f_{\h_{p,j,t}}(\y|\x_{p,j,t},\x_{b,t}) \\
    &\qquad\qquad+ \lambda_b \sum_t \rho_t\sum_{k=1}^K w_{b,k,t} \E \hat f_{\h_{b,k,t}}(\y|\x_{p,t},\x_{b,k,t})
    + o(\|\h\|^2) \\
  &\qquad= \lambda f(\y)
    + \lambda \sum_t\rho_t\frac{\mu_2(K)\h'\text{diag}(\Hcal_f(\y|\x_t))\h}{2}
    + o(\|\h\|^2) \\
    &\qquad\qquad+ \lambda_p \sum_t\rho_t\sum_{j=1}^J w_{p,j,t} f_{\h_{p,j,t}}(\y|\x_{p,j,t},\x_{b,t}) \\
    &\qquad\qquad+ \lambda_p\sum_t\rho_t\sum_{j=1}^Jw_{p,j,t}
      \frac{\mu_2(K)\h_{p,j,t}'\text{diag}(\Hcal_f(\y|\x_{p,j,t},\x_{b,t}))\h_{p,j,t}}{2} \\
    &\qquad\qquad+ \lambda_b \sum_t\rho_t\sum_{k=1}^K w_{b,k,t} f_{\h_{b,k,t}}(\y|\x_{p,t},\x_{b,k,t}) \\
     &\qquad\qquad+ \lambda_b\sum_t\rho_t\sum_{k=1}^K w_{b,k,t}
      \frac{\mu_2(K)\h_{b,k,t}'\text{diag}(\Hcal_f(\y|\x_{p,t},\x_{b,k,t}))\h_{b,k,t}}{2},
\end{align*}
and
\begin{align*}
  \Var(\hat f_{\lambdabf}(\y)) &= \Var\left(\lambda \hat f_\h(\y)
    + \lambda_p \hat f_{\text{sp}}(\y)
    + \lambda_b \hat f_{\text{sb}}(\y)\right) \\
  &= \lambda^2 \sum_t\rho_t\frac{R_{\x_t}(f)f(\y|\x_t)}{n\det(\Hbf)} + O\left(\frac{1}{n}\right)
    + \lambda_p^2\sum_t\rho_t \sum_{j=1}^J w_{p,j,t}^2 \Var \hat f_{\h_{p,j,t}}(\y|\x_{p,j,t},\x_{b,t}) \\
    &\qquad+ \lambda_b^2\sum_t\rho_t \sum_{k=1}^K w_{b,k,t}^2 \Var \hat f_{\h_{b,k,t}}(\y|\x_{p,t},\x_{b,k,t}) \\
  &= \lambda^2 \sum_t\rho_t \frac{R_{\x_t}(f)f(\y|\x_t)}{n\det(\Hbf)} + O\left(\frac{1}{n}\right)
    + \lambda_p^2 \sum_t\rho_t \sum_{j=1}^J w_{p,j,t}^2
      \frac{R_{\tilde \x_{p,j,t}}(f)f(\y|\x_{p,j,t},\x_{b,t})}{n_{p,j,t}\det(\Hbf_{p,j,t})} \\
    &\qquad+ \lambda_b^2 \sum_t\rho_t \sum_{k=1}^K w_{b,k,t}^2
      \frac{R_{\tilde\x_{b,k,t}}(f)f(\y|\x_{p,t},\x_{b,k,t})}{n_{b,k,t}\det(\Hbf_{b,k,t})},
\end{align*}

We will define $\tilde \x_{b,k,t} = (\x_{p,t}',\x_{b,k,t}')'$ and $\tilde \x_{p,j,t} = (\x_{p,j,t}',\x_{b,t}')'$ for notational convenience, and will additionally assume the following regularity approximations:
\begin{itemize}
\item[A4.] The quantities $\sum_{j=1}^Jw_{p,j,t}^2\|\x_t - \tilde \x_{p,j,t}\|^m$ and $\sum_{k=1}^Kw_{b,k,t}^2\|\x_t - \tilde \x_{b,k,t}\|^m$ are negligible, where $m = 2,4$.
\item[A5.] The quantities $\sum_{j=1}^Jw_{p,j,t}\left(\h_{p,j,t}'\text{diag}(\Hcal_f(\y|\x_{p,j,t},\x_{b,t}))\h_{p,j,t} - \h'\text{diag}(\Hcal_f(\y|\x_t))\h\right)$ and \\
$\sum_{k=1}^Kw_{b,k,t}\left(\h_{b,k,t}'\text{diag}(\Hcal_f(\y|\x_{p,t},\x_{b,k,t}))\h_{b,k,t} - \h'\text{diag}(\Hcal_f(\y|\x_t))\h\right)$ are negligible.
\end{itemize}
Approximation A4 is reasonable in our baseball application where there are many players similar enough to the players under study so that $\sum_{j=1}^Js_{p,j,t} > 1$ and $\sum_{k=1}^Ks_{b,k,t} > 1$ and $s_{p,j,t}\|\x_t-\tilde\x_{p,j,t}\|^m, s_{b,k,t}\|\x_t-\tilde\x_{b,k,t}\|^m \to 0$ as $\|\x_t-\tilde\x_{p,j,t}\|,\|\x_t-\tilde\x_{b,k,t}\| \to \infty$ for all integers $m$. Approximation A5 is reasonable by similar logic. Specification of $\beta = 2$ implies that
$\|\text{diag}(\Hcal_f(\y|\x_{p,t},\x_{b,k,t})) - \text{diag}(\Hcal_f(\y|\x_t))\| \leq \sqrt{d_p}L$ and
$\|\text{diag}(\Hcal_f(\y|\x_{p,j,t},\x_{b,t})) - \text{diag}(\Hcal_f(\y|\x_t))\| \leq \sqrt{d_b}L$
where $d_p$ and $d_b$ are, respectively, the dimension of $\x_{p,t}$ and $\x_{b,t}$. 
Let $\theta_{p,j,t} = n\det(\Hbf)/n_{p,j,t}\det(\Hbf_{p,j,t})$ and $\theta_{b,k,t} = n\det(\Hbf)/n_{b,k,t}\det(\Hbf_{b,k,t})$. With these specifications, we have that
\begin{align*}
  &\Var(\hat f_{\lambdabf}(\y)) - \Var(\hat f_\h(\y)) + O\left(\frac{1}{n}\right) \\
  &=(\lambda^2 - 1)\sum_t\rho_t\frac{R_{\x_t}(f)f(\y|\x_t)}{n\det(\Hbf)}
    + \lambda_p^2\sum_t\rho_t \sum_{j=1}^J \theta_{p,j,t}w_{p,j,t}^2
      \frac{R_{\tilde \x_{p,j,t}}(f)f(\y|\x_{p,j,t},\x_{b,t})}{n\det(\Hbf)} \\
    &\qquad+ \lambda_b^2 \sum_{k=1}^K \theta_{b,k,t}w_{b,k,t}^2
      \frac{R_{\tilde\x_{b,k,t}}(f)f(\y|\x_{p,t},\x_{b,k,t})}{n\det(\Hbf)} \\
  &\leq \sum_t\rho_t\left(\lambda^2 + \lambda_p^2 \sum_{j=1}^J \theta_{p,j,t}w_{p,j,t}^2
    + \lambda_b^2 \sum_{k=1}^K \theta_{b,k,t}w_{b,k,t}^2 - 1\right)
      \frac{R_{\x_t}(f)f(\y|\x_t)}{n\det(\Hbf)} \\
    &\qquad+ \lambda_p^2 \sum_t\rho_t\sum_{j=1}^J \theta_{p,j,t}w_{p,j,t}^2
      \left(\frac{R_{\x_t}(f)\|\x_t-\tilde \x_{p,j,t}\|^2
      + 2Lf(\y|\x_t)\|\x_t-\tilde \x_{p,j,t}\|^2 + 2L^2\|\x_t-\tilde \x_{p,j,t}\|^4}{n\det(\Hbf)}\right) \\
    &\qquad+ \lambda_b^2\sum_t\rho_t \sum_{k=1}^K \theta_{b,k,t}w_{b,k,t}^2
      \left(\frac{R_{\x_t}(f)\|\x_t-\tilde \x_{b,k,t}\|^2
      + 2Lf(\y|\x_t)\|\x_t-\tilde \x_{b,k,t}\|^2 + 2L^2\|\x_t-\tilde \x_{b,k,t}\|^4}{n\det(\Hbf)}\right).
\end{align*}
Assumption A4 and an identical lower bound argument implies that
$
  \Var(\hat f_{\lambdabf}(\y)) - \Var(\hat f_\h(\y))
$
is approximately bounded above by
$$
   \sum_t\rho_t\left(\lambda^2 + \lambda_p^2 \sum_{j=1}^J\theta_{p,j,t}w_{p,j,t}^2
      + \lambda_b^2 \sum_{k=1}^K \theta_{b,k,t}w_{b,k,t}^2 - 1\right)
      \frac{R_{\x_t}(f)f(\y|\x_t)}{n\det(\Hbf)} + O\left(\frac{1}{n}\right).
$$

We also have
\begin{align*}
  \text{Bias}(\hat f_\h(\y),f(\y))^2
    &= \left(\sum_t\rho_t\frac{\mu_2(K)\h'\text{diag}(\Hcal_f(\y|\x_t))\h}{2}
    + o(\|\h\|^2)\right)^2,
\end{align*}
and regularity approximations A4 and A5 yield
\begin{align*}
  &\text{Bias}(\hat f_{\lambdabf}(\y),f(\y))^2 = \left((\lambda - 1)\sum_t\rho_tf(\y|\x_t)
    + \lambda \sum_t\rho_t\frac{\mu_2(K)\h'\text{diag}(\Hcal_f(\y|\x_t))\h}{2}
      + o(\|\h\|^2)\right. \\
    &\qquad+ \lambda_p \sum_t\rho_t \sum_{j=1}^J w_{p,j,t} f(\y|\x_{p,j,t},\x_{b,t})
      + \lambda_p\sum_t\rho_t\sum_{j=1}^Jw_{p,j,t}
      \frac{\mu_2(K)\h_{p,j,t}'\text{diag}(\Hcal_f(\y|\x_{p,j,t},\x_{b,t}))\h_{p,j,t}}{2} \\
    &\qquad+ \left.\lambda_b\sum_t\rho_t \sum_{k=1}^K w_{b,k,t} f(\y|\x_{p,t},\x_{b,k,t})
      + \lambda_b\sum_t\rho_t\sum_{k=1}^K w_{b,k,t}
      \frac{\mu_2(K)\h_{b,k,t}'\text{diag}(\Hcal_f(\y|\x_{p,t},\x_{b,k,t}))\h_{b,k,t}}{2}\right)^2 \\
 &\leq \left((\lambda-1)\sum_t\rho_t f(\y|\x_t)
    +  \lambda \sum_t\rho_t \frac{\mu_2(K)\h'\text{diag}(\Hcal_f(\y|\x_t))\h}{2} + o(\|\h\|^2)\right. \\
    &\qquad+ \lambda_p \sum_t\rho_t\sum_{j=1}^J w_{p,j,t}(f(\y|\x_t) + L(-1)^z\|\x_t - \tilde\x_{p,j,t}\|^2)
      + \lambda_p \sum_t\rho_t \sum_{j=1}^Jw_{p,j,t}\frac{\mu_2(K)\h'\text{diag}(\Hcal_f(\y|\x_t))\h}{2} \\
    &\qquad+ \left.\lambda_b \sum_t\rho_t \sum_{k=1}^K w_{b,k,t}(f(\y|\x_t)
      + L(-1)^z\|\x_t - \tilde\x_{b,k,t}\|^2)
      + \lambda_b\sum_t\rho_t\sum_{k=1}^K w_{b,k,t}\frac{\mu_2(K)\h'\text{diag}(\Hcal_f(\y|\x_t))\h}{2}\right)^2 \\
 &\approx \left(\lambda_p \sum_t\rho_t\sum_{j=1}^J (-1)^z Lw_{p,j,t}\|\x_t - \tilde\x_{p,j,t}\|^2
   + \lambda_b \sum_t\rho_t\sum_{k=1}^K (-1)^z Lw_{b,k,t}\|\x_t - \tilde\x_{b,k,t}\|^2\right. \\
   &\qquad+ \left.\sum_t\rho_t\frac{\mu_2(K)\h'\text{diag}(\Hcal_f(\y|\x_t))\h}{2} + o(\|\h\|^2)\right)^2,
\end{align*}
where $z \in \{0,1\}$ is chosen to satisfy the above inequality. Putting these variance and bias results together without the lower order terms yields
\begin{align*}
  &MSE(\hat f_{\lambdabf}(\y),f(\y)) - MSE(\hat f_{\h}(\y),f(\y)) \\
    &\qquad\leq    \sum_t\rho_t\left(\lambda^2 + \lambda_p^2 \sum_{j=1}^J\theta_{p,j,t}w_{p,j,t}^2
      + \lambda_b^2 \sum_{k=1}^K \theta_{b,k,t}w_{b,k,t}^2 - 1\right)
      \frac{R_{\x_t}(f)f(\y|\x_t)}{n\det(\Hbf)} \\
   &\qquad+ \left(\lambda_p \sum_t\rho_t\sum_{j=1}^J (-1)^z Lw_{p,j,t}\|\x_t - \tilde\x_{p,j,t}\|^2
   + \lambda_b \sum_t\rho_t\sum_{k=1}^K (-1)^z Lw_{b,k,t}\|\x_t - \tilde\x_{b,k,t}\|^2\right. \\
   &\qquad+ \left.\sum_t\rho_t\frac{\mu_2(K)\h'\text{diag}(\Hcal_f(\y|\x_t))\h}{2}\right)^2
\end{align*}

This motivates the following choice of $\lambdabf$,
$$
  \lambda = \frac{\sqrt{n}}{\sqrt{n} + \sqrt{n_p} + \sqrt{n_b}}, \qquad
  \lambda_p = \frac{\sqrt{n_p}}{\sqrt{n} + \sqrt{n_p} + \sqrt{n_b}}, \qquad
  \lambda_b = \frac{\sqrt{n_b}}{\sqrt{n} + \sqrt{n_p} + \sqrt{n_b}},
$$
where $n_p = \sum_t\rho_t\sum_{j=1}^J s_{p,j,t}^2n_{p,j,t}$ and $n_b = \sum_t\rho_t\sum_{k=1}^K s_{b,k,t}^2n_{b,k,t}$. We will now develop intuition for these choices. First, notice that $\lambda_p,\lambda_b \to 0$ as
$\min_j(\|\x_t - \tilde\x_{p,j,t}\|), \min_k(\|\x_t - \tilde\x_{b,k,f}\|) \to \infty$ for all $t = 1,\ldots,n_{\text{type}}$. These cases correspond, to there being no similar pitchers or batters to the players under study. We turn attention to the bias terms, notice that
\begin{align*}
  &\lambda_p\sum_t\rho_t \sum_{j=1}^J (-1)^z Lw_{p,j,t}\|\x_t - \tilde\x_{p,j,t}\|^2 \\
    &\qquad= \frac{\sqrt{\sum_t\rho_t\sum_{j=1}^J s_{p,j,t}^2n_{p,j,t}}\left(\sum_t\rho_t
      \sum_{j=1}^J (-1)^z Lw_{p,j,t}\|\x_t - \tilde\x_{p,j,t}\|^2\right)}
      {\sqrt{n} + \sqrt{n_p} + \sqrt{n_b}} \longrightarrow 0,
\end{align*}
when there exists some $j'$ such that $\|\x_t - \tilde\x_{p,j',t}\| \to 0$ or $\min_j(\|\x_t - \tilde\x_{p,j,t}\|) \to \infty$ for each pitch type $t = 1,\ldots, n_{\text{type}}$. These cases correspond, respectively, to there being a few highly similar pitchers or there being no similar pitchers for each pitch thrown by the pitcher under study. Thus, the discrepancy in bias vanishes in the extreme cases. The same argument holds for batters. Now notice that
\begin{align*}
  \lambda_p^2 \sum_t\rho_t\sum_{j=1}^J\theta_{p,j,t}w_{p,j,t}^2
    &= \frac{\left(\sum_t\rho_t\sum_{j=1}^J s_{p,j,t}^2n_{p,j,t}\right)\left(\sum_t\rho_t\sum_{j=1}^J\theta_{p,j,t}w_{p,j,t}^2\right)}
      {(\sqrt{n} + \sqrt{n_p} + \sqrt{n_b})^2} \\
  & \longrightarrow
      \left\{\begin{array}{cl}
       0, & \min_{j,t}(\|\x_t - \tilde\x_{p,j,t}\|) \to \infty; \\
       \frac{\sum_t\rho_t n_{p,j_t,t}}
         {(\sqrt{n} + \sqrt{\sum_t\rho_tn_{p,j_t,t}} + \sqrt{n_b})^2}, & w_{p,j_t,t} \to 1, \; \text{for all} \; t = 1,\ldots,n_{\text{type}},
      \end{array}\right.
\end{align*}
under the specifications that $\theta_{p,j_t,t} = 1$. The same argument holds for batters under the specifications that $\theta_{b,k_t,t} = 1$. Therefore, when there is a pitcher $j_t$ and batter $k_t$ so that  $w_{p,j_t,t},w_{b,k_t,t} \to 1$ for each pitch type $t = 1,\ldots,n_{\text{type}}$, we have that
$$
  \sum_t\rho_t\left(\lambda^2 + \lambda_p^2 \sum_{j=1}^J \theta_{p,j}w_{p,j}^2
    + \lambda_b^2 \sum_{k=1}^K \theta_{b,k}w_{b,k}^2 - 1\right) \longrightarrow
    \frac{n + \sum_t\rho_tn_{p,j_t,t} + \sum_t\rho_tn_{b,k_t,t}}
      {(\sqrt{n}
        + \sqrt{\sum_t\rho_tn_{p,j_t,t}} + \sqrt{\sum_t\rho_tn_{b,k_t,t}})^2} - 1.
$$
Our choices of the elements of $\lambdabf$ will work well in the presence or absence of pitchers and batters that recover the traits of the players under study. Less is known about middle ground cases, especially when sample sizes are are small.

\end{document}